\documentstyle[preprint,pra,aps]{revtex}

\begin{document}

\draft

\tightenlines

\title{Non-relativistic quantum electrodynamics for strong laser-atom interaction}

\author{Marco Frasca}
\address{Via Erasmo Gattamelata, 3,
         00176 Roma (Italy)}

\date{\today}

\maketitle

\abstract{A formulation of quantum electrodynamics is given that applies
to atoms in a strong laser field by perturbation theory in a non-relativistic
regime. Dipole approximation is assumed.
With the current wavelengths, squeezing can be proved to be negligible and then,
just the linear term in the Hamiltonian can be retained. The dual Dyson series,
here discussed by referring it to the Birkhoff theorem
for singularly perturbed linear differential equations,
can be applied and a perturbation series obtained transforming the Hamiltonian
by a Pauli-Fierz transformation. But, if just few photons are
present high-order harmonics cannot be generated. So,
it is proven that odd high-order harmonics only appear when the laser field
is intense and one can substitute the creation and annihilation operators
by the square root of the mean number of photons taken to be huge, the field
retaining its coherency property as observed experimentally for harmonics.
In this case, the Hamiltonian for perturbation theory comes to the
Kramers-Henneberger form. From this Hamiltonian
it is shown that just odd harmonics of the laser frequency
contribute to the spectrum for a spherically
symmetric potential. This contribution is dipolar
when the free-electron quiver motion amplitude is larger
than the atomic radius. For a Coulomb potential
one has that the outer electron is periodically kicked, and so
a prove is given that the same should happen to Rydberg atoms
in intense microwave fields.
The distribution representing the kicking
has a Fourier series with just odd terms.
Using a modified Rayleigh-Schr\"odinger perturbation theory,
it is shown that under the
same condition of validity of the quiver motion amplitude
to atomic radius ratio,
the atomic wave function is only slightly modified by the laser field
due to the
way the energy levels rearrange themselves. This gives a prove
of stabilization in the limit of laser frequency going to infinity.
Then, perturbation
theory can be applied when the product of the laser frequency and the
square root of the ratio between the ionization energy
and the ponderomotive energy, that is the Keldysh parameter,
becomes smaller with respect to the shifted distance
between the energy levels of the atom.
}

\pacs{PACS: 32.80.-t, 42.50.Ct, 42.65.Ky, 05.45.+a}

\narrowtext

\section{Introduction}

The availability of powerful sources of laser light has permitted in recent years the
study of light-matter interaction in regimes where known approximations
fail. Indeed, a number of new phenomena has appeared as high-order harmonics of the
laser frequency, ionization with a number of photons well above the requested
threshold and possibly, there are also strong theoretical indications of existence
of stabilization, meaning by this that the ionization rate goes to zero by
increasing the intensity of the laser field \cite{kk1}. There has been a lot of theoretical work
following the appearance of those effects and, as a general approach, people
aims to develop models that account for the physics of strong-laser atom
interaction trying to support them both by numerical and experimental work.
So, a very succesful model for harmonic generation has been firstly put
forward in \cite{Cork}. This is the recollision model where is assumed that
the outer electron goes into continuum due to the laser field and here is
accelerated by the field, this step being described classically, then it
recombines with the core generating the harmonics. A quantum improved version,
understanding the emitted radiation as bremsstrahlung radiation due to the
electron approaching the core has been given in \cite{kk2}. This improvement
has the advantage that an explanation of the appearance solely of odd
harmonics of the laser frequency is given. Another quantum account of
the recollision model has been given in \cite{Huil}. To obtain
these quantum models a number of reasonable assumptions have been made as the
disregarding of intermediate states, the absence of resonances and so on, that
need to be justified. These quantum versions of the recollision model
prove to be a satisfactory explanation in all the experiments carried out
so far with harmonics. Then, any theory starting from quantum electrodynamics
has to cope with such a succesful understanding of the situation at hand.

Since the initial work by Kulander et al. \cite{kul} a lot of numerical work is
currently performed to understand strong laser-atom interaction. Anyhow,
although this work is highly interesting our aim in this paper is quite different.
That is, we are trying to rederive the recollision model from purely quantum
arguments without any other approximation than the schemes of current
experiments in high-intensity laser field require. Our conclusions concern the
justification needed to understand the theoretical models discussed in 
Ref.\cite{kk2,Huil}. In this light, we take for granted that numerical
computations support the recollision model as the experiments do.

After the pioneering work in \cite{mil}, a lot
of studies have also been carried out assuming a simple two-level model. So,
one of the open problems in this field of research is if such a simple
model really accounts for the physics of harmonic generation. In this
light, the general structure of the spectrum has been obtained
by Floquet theory and
a parallel with the recollision model has been given in \cite{kei,maq} for
the so called cut-off law, that is the rule to obtain the region of
the spectrum where the intensity of the harmonics goes rapidly down. It
should be said that such a model cannot account for ionization and so
appears somewhat rough at best.

In the light of the above approaches, in this paper we will try to change
the point of view beginning directly from quantum electrodynamics,
taking as a starting point our work in Ref.\cite{fra0} that
here is deepened and improved.
Indeed, modelling is an important way of understanding but,
if one is able to solve the equations of the full theory without
resorting to numerical methods, again a deep way to understand physics
is also given. So, our main aim is to study in depth what is the
physics of strong laser-atom interaction by directly solving the
Schr\"odinger equation. The theory we obtain is anyway enough
general to be possibly applicable to other physical situations
where strong electromagnetic fields play a role as e.g for Rydberg atoms.

It is interesting to note that, in view of using the dual Dyson series
presented in \cite{fra1} and discussed here by using the Birkhoff theorem \cite{ma},
the Pauli-Fierz transformation and its
classical counterpart, the Kramers-Hennerberger transformation, \cite{scul}
turn out to play a dominant role. Indeed, when unitary transformations
are applied, using probability amplitudes permits to extract physics
without worrying about. This is a standard approach e.g. in quantum
field theory for the interaction picture. Things do not change for
the dual Dyson series through the Pauli-Fierz transformation.

Some approximations to start with are needed, but we take them directly
from the experiments to tailor the Hamiltonian of quantum electrodynamics
to the kind of problems we want to discuss. So, we take the electromagnetic
field to be second quantized and the atom field in the
non-relativistic approximation. The quadratic term of
the vector potential in the Hamiltonian
is considered just to see, through a single
mode approximation, that taking it into account would shift the
frequency of the laser in the harmonics. So, the condition to neglect
it is also given. Finally, as the wavelengths of interest are much larger
than the atomic radius, the dipole
or long wavelength approximation is also taken. Then,
a Pauli-Fierz transformation is applied to study strong laser-atom
interaction. We will show that only if a large number of photons is
present and the amplitude of the quiver motion of the free-electron
is much larger than the atomic radius, a condition normally met in
this kind of experiments and a key approximation that we will assume
in this paper, than just odd harmonics are generated,
assuming the atomic potential to be spherically symmetric. Indeed, for a
large number of photons the Pauli-Fierz transformation reduces to
the Kramers-Henneberger transformation \cite{kh} as should be.

An interesting result obtained in this way is that, the atom appears
to be a kicked quantum system, being this result rigorously derived
from the Hamiltonian of quantum electrodynamics. The kicking
hypothesis has been assumed for Rydberg atoms in a microwave field
firstly in Ref.\cite{cas1} in order to explain the experimental results about
ionization. Indeed, kicking can mean localization as a counterpart of classical
chaos \cite{cas2}. This is one of the main results of this paper
pointing toward the merging of such different fields of research.

A kicked Hamiltonian can in principle be solved exactly.
Anyhow,
we want to study an atom in a strong laser field doing perturbation theory.
So, we will obtain a perturbation series having the product $\omega\gamma$
between the
Keldysh parameter $\gamma=\sqrt{\frac{I_B}{2U_p}}$, being
$I_B$ the ionization
energy and $U_p=\frac{e^2E^2}{4m\omega^2}$ the ponderomotive energy
for an electric field $E$,
and the laser frequency $\omega$
smaller than the distances, properly shifted by the laser field,
between the energy levels. This accomplish the task to obtain a perturbation
theory dual to the standard time-dependent theory applied to an atom
in a weak electromagnetic field. Using this approach,
we are able to show that the rate of above threshold ionization
determines the duration of the harmonics in the spectrum and then,
also the form of the spectrum itself by Fourier transform.
Beside, if a rough model for the outer
electron in the field of the rest of the atom is taken having a Coulomb
form with a $Z_{eff}$ for the atomic number to give the correct ionization
energy, it can be seen that the duration of harmonics agrees fairly well
with experimental results as given e.g. in \cite{hb}. This simply means
to scale all the formulas for hydrogen-like atoms by ionization potential,
ponderomotive energy and laser frequency.

The applicability
of perturbation theory is indeed possible as,
through a modified Rayleigh-Schr\"odinger
perturbation theory one can show that the energy levels of the atom are
shifted in such a way to change very little the wave function
giving a prove of stabilization in the limit of
laser frequency increasingly large. This is
what we call ``rigidity'' of the wave function, very similar to the
behavior of a superconductor due to the presence of the
binding energy of the Cooper pair. This should give a hint, in the
framework of quantum chaos, to study the change of
statistics of such energy levels, possibly by some numerical work.

Let us finally point out that a number of methods have been devised
starting from Volkov states, that are the solution of the Schr\"odinger
equation for a free particle in a plane wave. The prototype of these
approaches is given in Ref.\cite{kel}. Anyhow, our point of view
assumes different asymptotic states that are naturally derived
from dual Dyson series as an entangled state between the radiation
and the atomic state: This is just the leading order approximation.
Then, e.g. no hypothesys a priori on the atomic
potential being short ranged is needed. On the other side, no general
proof is known of Volkov states being the proper asymptotic states
for perturbation theory in a strong electromagnetic field. Beside,
it appears very difficult to understand why just odd harmonics
are experimentally observed as this approach keeps both odd and
even harmonics \cite{bec}. Anyhow, an improved version has been recently given
in Ref.\cite{gao} by the Volkov states in a second quantized field
giving just odd harmonics. This approach satisfy the principle of duality
in perturbation theory but what is changed is the initial state used to
obtain the perturbation series and is generally presented as a non-perturbative
method through the results for the computation of the amplitudes given in
Ref.\cite{kel}.

The paper is so structured. In sec.\ref{dua} we give a brief presentation of
duality for the Schr\"odinger equation and discuss the Birkhoff theorem
using the dual Dyson series. In sec.\ref{for} we give the Hamiltonian
formulation of quantum electrodynamics to start with and show why
the quadratic term is negligible. Then we perform a Pauli-Fierz
transformation to obtain the dual Hamiltonian and show how it
reduces to the Kramers-Henneberger form. A discussion of
the two-level approximation with respect to the full theory
is also given. In sec.\ref{kic} we apply
the approximation of large amplitude of the quiver motion in
the laser field with the respect to the atomic radius to show
how kicking arises and why the atomic wave function turns out
to be rigid in the sense given above. In sec.\ref{per} perturbation
theory is done in the tunnelling regime deriving the
rate of ionization and consequently the spectrum
of the harmonics. Finally, in sec.\ref{end} the conclusions are given.

\section{Duality and Birkhoff Theorem}
\label{dua}

The duality principle in perturbation theory,
when applied to the Schr\"odinger equation
written as (here and in the following $\hbar=c=1$)
\begin{equation}
    (H_0+H_1(t))|\psi\rangle = i\frac{\partial|\psi\rangle}{\partial t} \label{eq:sch}
\end{equation}
states that, by choosing $H_0$ as unperturbed Hamiltonian, the
perturbation series has a development parameter exactly the inverse
of the series obtained by taking $H_1$ as unperturbed Hamiltonian.
Formally, this latter case means that we are multiplying $H_1$ by
an ordering parameter $\lambda$ going to infinity. Duality principle is
enough, as shown in Ref.\cite{fra1,fra2}, to prove that the dual to
the Dyson series is just the adiabatic approximation and its higher
order corrections. Indeed, from the theory of singular perturbation
we can rederive a similar result
for the equations of probability amplitudes by the Birkhoff theorem
\cite{ma}, where ``singular'' means that the small parameter multiplies
the higher derivative in the equation. In this way, a dual Dyson series can be
obtained also for the probability amplitudes. In fact, the dual perturbation
series is obtained by solving at the leading order the equation
\begin{equation}
    \lambda H_1(t)|\psi^{(0)}\rangle = i\frac{\partial|\psi^{(0)}\rangle}{\partial t}
\end{equation}
taking $\lambda\rightarrow\infty$. But this can be stated as a singular
perturbation problem. Indeed, using the eigenstates of the
unperturbed part $H_0|n>=E_n|n>$ and putting $|\psi(t)>=\sum_ne^{-iE_nt}a_n(t)|n>$,
one has for eq.(\ref{eq:sch}) the system of linear differential equations for the
amplitudes
\begin{equation}
    i\epsilon\frac{da_m(t)}{dt}=\sum_n e^{-i(E_n-E_m)t}
    \langle m|H_1(t)|n\rangle a_n(t) \label{eq:bk}
\end{equation}
being now $\epsilon=\frac{1}{\lambda}$. If the Hamiltonian $H_0$ has a finite set of
$N$ eigenstates and eigenvalues and considering the amplitudes $a_m(t)$
as elements of a vector ${\bf a}(t)$, we can
apply a result obtained by Birkhoff on 1908
in singular perturbation theory \cite{ma}, that states that the
equation (\ref{eq:bk}) has a fundamental set of solutions ${\bf b}_j(t)$
with $1\le j\le N$ if the eigenvalues $\lambda_j(t)$ are distinct, given by
\begin{equation}
    b_{kj}(t)=\exp\left[-i\int_{t_0}^t\lambda_j(t')dt'\right]
              \exp[i\gamma_j(t)]u_{kj}(t)+O(\epsilon)
\end{equation}
$k$-th element of the vector ${\bf b}_j(t)$,
$\gamma_j(t)=\int_{t_0}^t{\bf u}_j^*(t')i\frac{d}{dt'}{\bf u}_j(t')$
the geometrical part of the phase, ${\bf u}_j(t)$
the eigenstates of the matrix having elements $A_{mn}=e^{-i(E_n-E_m)t}
\langle m|H_1(t)|n\rangle$. The condition of no crossing of eigenvalues $\lambda_j(t)$
is also required. The final leading order approximation is then written as
\begin{equation}
    {\bf a}(t)=\sum_{j=1}^N\alpha_j{\bf b}_j(t)
\end{equation}
with the coefficient $\alpha_j$ given by the initial conditions ${\bf a}(0)$.
Again, we realizes that formally this is the adiabatic approximation
applied to the operator $e^{iH_0t}H_1(t)e^{-iH_0t}$.
Then, we are arrived at similar conclusions as in Ref.\cite{fra2} for dressed states where,
instead, duality principle was used, in the limit $\lambda\rightarrow\infty$.
But, the use of duality principle permits to get rid of the limitation on the spectrum
of $H_0$ being bounded to $N$ eigenstates and eigenvalues and then, it appears as a
more general tool to treat also this kind of problems. So, we relax the condition on
the spectrum of $H_0$ required by the Birkhoff theorem, as duality permits
to give an alternative way to derive it.

Indeed, one can build the dual Dyson series by also doing an unitary transformation that removes
the perturbation $H_1(t)$ in eq.(\ref{eq:sch}). This defines a dual interaction picture.
In fact, one can see that the unitary transformation operator
must be a solution of
\begin{equation}
    \lambda H_1(t)U_F(t) = i\frac{\partial U_F(t)}{\partial t}
\end{equation}
and so $U_F(t)$ turns out to be the evolution operator of the adiabatic approximation
in the limit $\lambda\rightarrow\infty$, proving the equivalence. But, this equation
can also be solved exactly in some cases, without resorting to the adiabatic approximation
giving the series
\begin{equation}
|\psi(t)\rangle=U_F(t){\cal T}\exp\left[-i\int_{t_0}^tdt'U_F^\dagger(t')H_0U_F(t')\right]|\psi(t_0)\rangle
\label{eq:dpt}
\end{equation}
with ${\cal T}$ the time-ordering operator.

\section{Non-relativistic Formulation of Quantum Electrodynamics}
\label{for}

\subsection{Full Hamiltonian and the Quadratic Term}

The starting assumptions about the formulation of quantum electrodynamics that we
need are the following. Firstly, we takes a non-relativistic approximation and
neglect spin effects. Secondly, we cut-off the wavelengths limiting the analysis
to the long ones that is, we take the dipole approximation: currently, this is in
agreement with the experimental results for harmonic generation. Finally, we do
second quantization on the electromagnetic field as we want to understand the
properties of the scattered light in experiments with strong fields. So, we
write the Hamiltonian in the Coulomb gauge \cite{scul}
(here and in the following $\hbar=c=1$)
\begin{equation}
    H=\sum_{\lambda,{\bf k}}\omega_{\bf k}a^\dagger_\lambda({\bf k})a_\lambda({\bf k})
      +\frac{{\bf p}^2}{2m}+V({\bf x})+\frac{e}{m}{\bf A}{\bf p}+\frac{e^2}{2m}{\bf A}^2 \label{eq:h1}
\end{equation}
with $\lambda=1,2$ meaning a sum over polarizations, $V({\bf x})$ the atomic potential,
$a^\dagger_\lambda({\bf k}),a_\lambda({\bf k})$ the creation and annihilation operators
for the mode ${\bf k}$ with polarization $\lambda$. So, we have
\begin{equation}
    {\bf A}=\sum_{\lambda,{\bf k}}g_{\bf k}
    (\mbox{\boldmath$\epsilon$}^*_\lambda a^\dagger_\lambda({\bf k})+
     \mbox{\boldmath$\epsilon$}_\lambda a_\lambda({\bf k}))
\end{equation}
being $\mbox{\boldmath$\epsilon$}_\lambda$ the complex polarization vector,
$g_{\bf k}=\sqrt{\frac{2\pi}{V\omega_{\bf k}}}$ and a normalization in a
box of volume $V$ is assumed everywhere.

To see how much relevant is the quadratic term, we refer to the experimental
result that just harmonics of the laser frequency are observed. Then,
specializing the above Hamiltonian to a single mode \cite{scul}, we derive
the Heisenberg motion equation for the creation and annihilation operators.
So, we take
\begin{equation}
    H=\omega a^\dagger a
      +\frac{{\bf p}^2}{2m}+V({\bf x})+
      \frac{e}{m}{\bf A}{\bf p}+\frac{e^2}{2m}{\bf A}^2.
\end{equation}
Assuming a linear polarization, that is
\begin{equation}
    {\bf A}=\sqrt{\frac{2\pi}{V\omega}}\mbox{\boldmath$\epsilon$}(a^\dagger + a)
\end{equation}
we can use the same argument given in Ref.\cite{fra3}
that the laser frequency should be shifted
by squeezing. Indeed, one has the Heisenberg equations for $a$ and $a^\dagger$
\begin{eqnarray}
   \frac{da}{dt}&=&-i\omega a -i\frac{e^2g^2}{m}(a+a^\dagger)
   -i\frac{eg}{m}\mbox{\boldmath $\epsilon$}{\bf p} \nonumber \\
   \frac{da^\dagger}{dt}&=&i\omega a +i\frac{e^2g^2}{m}(a+a^\dagger)
   +i\frac{eg}{m}\mbox{\boldmath $\epsilon$}{\bf p}
\end{eqnarray}
and without going into details of computation,
by duality, as a first approximation,
one neglects the atomic Hamiltonian and then,
managing ${\bf p}$ as a c-number, the
time dependence of the creation and annihilation operators is harmonic
with frequency $\Omega=\sqrt{\omega^2+\frac{4\pi e^2}{mV}}$ and not just
$\omega$ as should be required by the experimental results. So, squeezing
terms can be neglected otherwise a shifted frequency of the harmonics
would be observed \cite{fra3}. This effect should be more pronounced
as greater is the order of the harmonic.
But, considering this shift with current frequencies and
gas densities really negligible, the quadratic
term of the field in the Hamiltonian (\ref{eq:h1}) can be systematically
neglected giving the final Hamiltonian
\begin{equation}
    H=\sum_{\lambda,{\bf k}}\omega_{\bf k}a^\dagger_\lambda({\bf k})a_\lambda({\bf k})
      +\frac{{\bf p}^2}{2m}+V({\bf x})+\frac{e}{m}\sum_{\lambda,{\bf k}}g_{\bf k}
    (\mbox{\boldmath$\epsilon$}^*_\lambda a^\dagger_\lambda({\bf k})+
     \mbox{\boldmath$\epsilon$}_\lambda a_\lambda({\bf k})){\bf p}. \label{eq:hf}
\end{equation}
This will be the starting point for our further analysis.

\subsection{Pauli-Fierz Transformation and Kramers-Henneberger Hamiltonian}

We now try to approach the Hamiltonian (\ref{eq:hf}) using duality principle,
that is, we try to compute the dual Dyson series. So, one can remove the
perturbation by a Pauli-Fierz transformation given by \cite{scul}
\begin{equation}
    U_{PF}=\exp\left[\sum_{{\bf k},\lambda}\left(\beta_\lambda^*({\bf k})a_\lambda({\bf k})
           -\beta_\lambda({\bf k})a_\lambda^\dagger({\bf k})\right)\right]
\end{equation}
where one has
\begin{equation}
    \beta_\lambda({\bf k})=\frac{e}{m\omega_{\bf k}}g_{\bf k}\mbox{\boldmath $\epsilon$}^*\cdot{\bf p}
\end{equation}
and obtains the Hamiltonian \cite{scul}
\begin{equation}
    H_{PF}=\sum_{{\bf k},\lambda}\omega_{\bf k}a^\dagger_\lambda({\bf k})a_\lambda({\bf k})+
           \frac{{\bf p}^2}{2m^*}+V\left[{\bf x}-i\sum_{{\bf k},\lambda}\left(\beta_\lambda^*({\bf k})a_\lambda({\bf k})
           -\beta_\lambda({\bf k})a_\lambda^\dagger({\bf k})\right)\right]
\end{equation}
being $m^*$ the renormalized mass due to the field. Here we try another way, to agree with duality in
perturbation theory. That is, firstly we consider the Hamiltonian (\ref{eq:hf}) in the interaction picture
giving
\begin{equation}
    H_I=\frac{{\bf p}^2}{2m}+V({\bf x})+\frac{e}{m}\sum_{\lambda,{\bf k}}g_{\bf k}
    \left[\mbox{\boldmath$\epsilon$}^*_\lambda a^\dagger_\lambda({\bf k})e^{i\omega_{\bf k}t}+
     \mbox{\boldmath$\epsilon$}_\lambda a_\lambda({\bf k})e^{-i\omega_{\bf k}t}\right]\cdot{\bf p}.
\end{equation}
Then, we solve the leading order equation
\begin{equation}
     \frac{e}{m}\sum_{\lambda,{\bf k}}g_{\bf k}
     \left[\mbox{\boldmath$\epsilon$}^*_\lambda a^\dagger_\lambda({\bf k})e^{i\omega_{\bf k}t}+
     \mbox{\boldmath$\epsilon$}_\lambda a_\lambda({\bf k})e^{-i\omega_{\bf k}t}\right]\cdot{\bf p}U_F(t)=
     i\frac{\partial U_F(t)}{\partial t}.
\end{equation}
The solution is standard and can be written as
\begin{equation}
    U_F(t)=\exp\left[-\sum_{{\bf k},\lambda}\gamma_{{\bf k},\lambda}(t)
           (\mbox{\boldmath $\epsilon$}_\lambda^*\cdot{\bf p})
           a_\lambda^\dagger({\bf k})\right]
           \exp\left[\sum_{{\bf k},\lambda}\gamma_{{\bf k},\lambda}^*(t)
           (\mbox{\boldmath $\epsilon$}_\lambda\cdot{\bf p})
           a_\lambda({\bf k})\right]
           \exp\left[\sum_{{\bf k},\lambda}\alpha_{{\bf k},\lambda}(t)
           (\mbox{\boldmath $\epsilon$}_\lambda^*\cdot{\bf p})(\mbox{\boldmath $\epsilon$}_\lambda\cdot{\bf p})\right]
\end{equation}
with
\begin{equation}
    \gamma_{{\bf k},\lambda}=\frac{e}{m}\sqrt{\frac{2\pi}{\omega_{\bf k}V}}
                             \frac{e^{i\omega_{\bf k}t}-1}{\omega_{\bf k}}
\end{equation}
and
\begin{equation}
    \alpha_{{\bf k},\lambda}=-\frac{1}{2}|\gamma_{{\bf k},\lambda}(t)|^2.
\end{equation}
So, we have found in this way a time-dependent version of the above Pauli-Fierz transformation. But now,
by eq.(\ref{eq:dpt}) we have a dual Dyson series for quantum electrodynamics.
From this, one has e.g. that the leading order
wave-function in the interaction picture,
if we assume the atom in the ground state, is given by
\begin{equation}
    |\psi^{(0)}(t)\rangle_I=U_F(t)|1s\rangle|\alpha\rangle
\end{equation}
being $|1s\rangle$ the atomic state and
$|\alpha\rangle$ the initial state of the laser field
described by a coherent state.
As it should be expected this is
an entangled state between the atomic state and the field
state \cite{fra3}. It is easy to realize, using
the momentum representation for the atomic state,
that the scattered light is indeed coherent and preserves
the property of the initial state for the field.

On this basis,
we expect the light of the harmonics to be
coherent as well \cite{fra3}. In fact, in order to evaluate
higher orders in the dual Dyson series we
rewrite the Pauli-Fierz transformation as
\begin{equation}
    U_F(t)=\exp\left[\sum_{{\bf k},\lambda}\left(\gamma_{{\bf k},\lambda}^*(t)
           (\mbox{\boldmath $\epsilon$}_\lambda\cdot{\bf p})
           a_\lambda({\bf k})-\gamma_{{\bf k},\lambda}(t)
           (\mbox{\boldmath $\epsilon$}_\lambda^*\cdot{\bf p})
           a_\lambda^\dagger({\bf k})\right)\right],
\end{equation}
obtaining the transformed Hamiltonian for higher order computations
\begin{equation}
    H_F=U_F^\dagger(t)\left(\frac{{\bf p}^2}{2m}+V({\bf x})\right)U_F(t)
       =\frac{{\bf p}^2}{2m}+V\left[{\bf x}+{\bf X}(t)\right] \label{eq:hf1}
\end{equation}
being
\begin{equation}
    {\bf X}(t)=-i\sum_{{\bf k},\lambda}\left(\gamma_{{\bf k},\lambda}^*(t)
           \mbox{\boldmath $\epsilon$}_\lambda
           a_\lambda({\bf k})-\gamma_{{\bf k},\lambda}(t)
           \mbox{\boldmath $\epsilon$}_\lambda^*
           a_\lambda^\dagger({\bf k})\right).
\end{equation}

Now, we specialize the above construction assuming that initially
the atom is exposed to an intense laser field.
It is well-known that the creation and annihilation operators can be
expressed through the number operator
$n_{{\bf k},\lambda}=a^\dagger_{{\bf k},\lambda}a_{{\bf k},\lambda}$
and a phase operator by the
Susskind and Glogower \cite{sg} or the Pegg and
Barnett construction \cite{pb}. But, the distribution probability of the
phase can be made to coincide in such a way to give for a coherent
state a definite phase having quantum fluctuactions that go to zero
for a very large mean number of photon, that is the case
of the intensity of the laser field of interest. Then,
one can keep just one mode and operate the substitution \cite{hak}
\begin{equation}
    a_{{\bf k},\lambda}\rightarrow
    \sqrt{n_{{\bf k},\lambda}}e^{i\phi_{{\bf k},\lambda}}.
\end{equation}
The Hamiltonian $H_F$ is then reduced to the Kramers-Henneberger form
\cite{kh} for a classical field that is,
\begin{equation}
    H_{HK}=\frac{{\bf p}^2}{2m}+
    V\left[{\bf x}-\lambda_L\mbox{\boldmath $\epsilon$}(\sin(\omega t -\phi)
           +\sin(\phi))\right] \label{eq:hkh}
\end{equation}
where a linear polarization has been assumed,
all subscripts have been dropped and $\lambda_L$,
the free-electron maximum excursion in a monochromatic field, can be
rewritten as $\lambda_L=\frac{eE}{m\omega^2}$,
being $E$ the intensity of the laser field. The above
Hamiltonian can be obtained by an unitary transformation,
the Kramers-Henneberger transformation, for
an atom in a monochromatic classical field. Then, we simply
neglect quantum fluctuaction beyond the leading order due to the
characteristic of the electromagnetic field.
The relevance of this limit
for a coherent state is that it makes consistent the arguments that
follow about the properties of the harmonics. We just note that
the advantage of such a derivation is that it gives us the
coherence property of the scattered laser light in atomic
scattering experiments, and permits to understand the way
from a quantum electrodynamics formulation to a
classical Hamiltonian that fully accounts for the situation at hand.

As a general approach,
in agreement with the duality principle in perturbation theory \cite{fra1},
we see that to study
strong laser-atom interaction we apply unitary transformations to remove
the big part of the
Hamiltonian that in this case is due to the field.
We now apply the same idea also to the two-level
model going to a dual interaction picture.

\subsection{Two-level Approximation}

The two-level model generally adopted in the study of harmonic generation
can be cast in the form (e.g. \cite{maq})
\begin{equation}
    H=\frac{\Delta}{2}\sigma_3+\Omega\sigma_1\cos(\omega t) \label{eq:tlm}
\end{equation}
being $\Delta$ the distance between the two-level, $\Omega$ the strength of the field, $\omega$ the
frequency of the laser field and $\sigma_1$, $\sigma_3$ the Pauli matrices. The first point to note
is that, in order to be in agreement with the recollision model, one of the two levels should be in
the continuum or, at best, $\frac{\Delta}{2}$ is the ionization energy $I_B$ \cite{kei}. In this way,
one can rederive the well-known cut-off law for the frequency of the maximum harmonic $I_B+3U_p$, being
$U_p=\frac{e^2E^2}{4m\omega^2}$ the ponderomotive energy that in this model should be proportional to
$\Omega$ as pointed out in Ref.\cite{kei}. But, in this model the continuum is missing and then,
ionization cannot be described. This means to lose any connection between ionization and harmonic
generation that, as we will see, plays a role in the duration of the harmonics itself. Finally,
let us apply a Pauli-Fierz transformation (or Kramers-Henneberger transformation) by removing the
oscillating term as
\begin{equation}
    U_F(t)=\exp\left[-i\sigma_1\frac{\Omega}{\omega}\sin(\omega t)\right]
\end{equation}
to obtain the transformed Hamiltonian
\begin{equation}
    H_F=\frac{\Delta}{2}J_0\left(\frac{2\Omega}{\omega}\right)+
        \Delta\sigma_3\sum_{n=1}^{\infty}J_{2n}\left(\frac{2\Omega}{\omega}\right)
        \cos(2n\omega t)+
        \Delta\sigma_2\sum_{n=0}^{\infty}J_{2n+1}\left(\frac{2\Omega}{\omega}\right)
        \sin((2n+1)\omega t).
\end{equation}
From this Hamiltonian is quite easy to realize that both odd and even harmonics play the
same role and the only way to make ones or the other appear is to choose proper initial
conditions. To show that things are really in this way, we turn back to Hamiltonian
(\ref{eq:tlm}) and apply the Birkhoff theorem to rederive the spectrum
already obtained in Ref.\cite{fra2},
given in a general form by the Floquet theory in Ref.\cite{maq}.

So, we have to solve the equation
\begin{equation}
\left[\frac{\Delta}{2}\sigma_3+\Omega\sigma_1\cos(\omega t)\right]
|\psi(t)\rangle=i\frac{d|\psi(t)\rangle}{dt} \label{eq:stlm}
\end{equation}
that in the interaction picture becomes the system of equations
\begin{eqnarray}
    i\frac{d}{dt}\left(\begin{array}{c} a_1(t) \\ a_2(t) \end{array}\right)=
    \left(\begin{array}{clcr} 0 & \Omega e^{-i\Delta t}\cos(\omega t) \\
                              \Omega e^{i\Delta t}\cos(\omega t) & 0
          \end{array}\right)\left(\begin{array}{c} a_1(t) \\ a_2(t) \end{array}\right) \label{eq:amp}
\end{eqnarray}
to which the Birkhoff theorem can be applied. It should be pointed out that
this approximation is
consistent with $\Delta\ll\Omega$ and we take $\omega\ll\Omega,\Delta$,
in agreement with the identifications of Ref.\cite{kei}
and in the way the model
should compare with experiments. Otherwise, the Birkhoff theorem should be
applied differently. Under these conditions,
we show that even harmonics can be present
depending on the initial conditions.
These harmonics are known as hyper-Raman lines
in the current literature \cite{maq,per}.

Firstly, we note that eq.(\ref{eq:amp}) is the same as eq.(\ref{eq:stlm})
when use has been made of the eigenstates of $\sigma_3$.
Then, we can simply use the results of Ref.\cite{fra2} about
the dressed states of Hamiltonian (\ref{eq:tlm}) and obtain
the Birkhoff basis as
\begin{equation}
    {\bf b}_1(t)=e^{-i\frac{\Delta}{2}t}e^{i\frac{\Omega}{\omega}
                  \sin(\omega t)}\frac{1}{\sqrt{2}}
                  \left(
                  \begin{array}{c} 1 \\ -e^{-i\Delta t} \end{array}
                  \right)
\end{equation}
and
\begin{equation}
    {\bf b}_2(t)=e^{i\frac{\Delta}{2}t}e^{-i\frac{\Omega}{\omega}
                  \sin(\omega t)}\frac{1}{\sqrt{2}}
                  \left(
                  \begin{array}{c} e^{i\Delta t} \\ 1 \end{array}
                  \right).
\end{equation}
Then, we conclude that the Birkhoff theorem gives the same results obtained
by other approaches in Ref.\cite{fra2,fra4}. The spectrum is then given by
\cite{fra2}
\begin{eqnarray}
    \langle\psi(t)|\sigma_1|\psi(t)\rangle&\approx&
    a_2(0)a_1^*(0)e^{-i\Delta_Rt}+a_2^*(0)a_1(0)e^{i\Delta_Rt} \nonumber \\
        &+&(|a_1(0)|^2-|a_2(0)|^2)\Delta
           \sum_{n=0}^{\infty}
           J_{2n+1}\left(\frac{2\Omega}{\omega}\right)
           \frac{\cos((2n+1)\omega t)-1}
           {(n+\frac{1}{2})\omega} \nonumber \\
       &+&i(a_2^*(0)a_1(0)e^{i\Delta_Rt}-a_2(0)a_1^*(0)e^{-i\Delta_Rt})
           \Delta\sum_{n=1}^{\infty}
           J_{2n}\left(\frac{2\Omega}{\omega}\right)
           \frac{\sin(2n\omega t)}{n\omega}
\end{eqnarray}
with $\Delta_R=\Delta J_0\left(\frac{2\Omega}{\omega}\right)$ the
renormalized level separation. This form of the spectrum is
in agreement with the one derived by the Floquet theory as given
in \cite{maq}. In the range of validity of this
approximation as stated above, this proves the assertion that the even
harmonics should be considered on the same ground as the odd ones
for the two-level model. As we are going to show, this makes the
applicability of this approximation to harmonic generation
somewhat unappropriate as the appearance of just odd harmonics in
the experiments has a deep physical meaning that here is overlooked.
In fact, the two-level model relies just on the initial conditions to
select the properties of the spectrum.

\section{Properties of Atoms in Intense Laser Fields}
\label{kic}
\subsection{Odd Harmonics and Kicking for an Electron in a Coulomb Field}

We now turn our attention to the classical limit of the Pauli-Fierz
Hamiltonian that is, the Kramers-Henneberger Hamiltonian that here
we rewrite assuming the phase $\phi=0$ obtaining
\begin{equation}
    H_{HK}=\frac{{\bf p}^2}{2m}+V\left[{\bf x}
           -\lambda_L\mbox{\boldmath $\epsilon$}\sin(\omega t)\right].
           \label{eq:hkhf}
\end{equation}
The potential has the Fourier series \cite{kk1}
\begin{equation}
    V\left[{\bf x}
           -\lambda_L\mbox{\boldmath $\epsilon$}\sin(\omega t)\right]
           =\sum_{k=0}^{\infty}v_k({\bf x})i^k
           \left[e^{ik\omega t}+(-1)^ne^{-ik\omega t}\right] \label{eq:vt}
\end{equation}
where
\begin{equation}
    v_k({\bf x})=\int_{-1}^1\frac{dx'}{\pi}
    V({\bf x}-\mbox{\boldmath $\epsilon$}\lambda_L x')
    \frac{T_k(x')}{\sqrt{1-x'^2}}
\end{equation}
and being $T_k(x)$ the $k$-th Chebyshev polynomial of first kind.
In this section we study the terms having $k>1$, deserving the analisys of
the static contribution with $k=0$, i.e. the dressed potential, for the
next section.

We now take the unperturbed potential $V({\bf x})$ to be spherical
symmetric. It easily realized that one has different parity for
even components being $v_{2n}(-{\bf x})=v_{2n}({\bf x})$, and
odd components being $v_{2n+1}(-{\bf x})=-v_{2n+1}({\bf x})$. But, the
unperturbed part of the Hamiltonian is invariant by parity transformation
while, the odd time-dependent part breaks this simmetry.
So, by perturbation theory
one has at first order $\langle\psi(t)|x|\psi(t)\rangle=
\langle\psi^{(0)}(t)|x|\psi^{(1)}(t)\rangle +c.c.$ being
$|\psi^{(0)}\rangle$ the initial atomic state such that
$\langle\psi^{(0)}(t)|x|\psi^{(0)}(t)\rangle=0$, and the only way not to
obtain a null spectrum through perturbation theory is from the odd components.
This result turns out to be in agreement with the general one obtained
through Floquet states in Ref.\cite{flo}.

To apply the above scheme to
the Coulomb potential one has to consider the integrals
\begin{equation}
    v_k({\bf x})=-\frac{Ze^2}{\lambda_L}\int_{-1}^1\frac{dx'}{\pi}
    \frac{1}{\sqrt{\left(\frac{\bf x}{\lambda_L}- x'\right)^2
    +\left(\frac{\bf y}{\lambda_L}\right)^2
    +\left(\frac{\bf z}{\lambda_L}\right)^2}}
    \frac{T_k(x')}{\sqrt{1-x'^2}} \label{eq:vkc}
\end{equation}
but we have to study them around the origin of coordinates where the
Coulomb potential turns out to be singular. Indeed, at small distances
the theory given above has its shortcomings due to the dipole approximation
and the neglecting of the relativity. Then, some regularization is needed
for the above integrals.
So, let us study the integrals (\ref{eq:vkc}) at the point $y=z=a_0$,
introducing in this way the cut-off $\eta=\frac{a_0}{\lambda_L}\ll 1$,
and we are reduced to the analysis of the leading term
\begin{equation}
    v_k({\bf x})=-\frac{Ze^2}{\lambda_L}\int_{-1}^1
    \frac{dx'}{\pi}
    \frac{1}{\sqrt{\left(\frac{\bf x}{\lambda_L}- x'\right)^2
    +2\eta^2}}
    \frac{T_k(x')}{\sqrt{1-x'^2}}
\end{equation}
very near the origin.
These integrals turn out to be zero for odd Chebyshev polynomials
when one takes $x=0$ and, for even Chebyshev polynomials one gets
an extremum such that the integral is bounded. The extremum depends
on $\eta$ but increases slowly. Then, we can remove the even
terms, after we take $v_{2n}$ at their values at the origin, by an unitary
transformation. The same cannot be done for odd terms as they go to
zero at the origin. These integrals can be studied by a Taylor series
at $x=0$ giving for the dipolar term
\begin{equation}
    v_k^{dip}({\bf x})=-\frac{Ze^2}{\lambda_L}\int_{-1}^1
    \frac{dx'}{\pi}
    \frac{1}{(x'^2
    +2\eta^2)^\frac{3}{2}}
    \frac{T_k(x')x'}{\sqrt{1-x'^2}}\frac{x}{\lambda_L}
\end{equation}
and the integral gives the cut-off dependent approximation
\begin{equation}
    \int_{-1}^1
    \frac{dx'}{\pi}
    \frac{1}{(x'^2
    +2\eta^2)^\frac{3}{2}}
    \frac{T_k(x')x'}{\sqrt{1-x'^2}}\approx
    -\frac{2}{\pi}\log(\eta)(2n+1)(-1)^n.
\end{equation}
that has a logarithmic divergence for the cut-off $\eta$ going to zero.
If we intrdouce the ratio $\frac{a_B}{\lambda_L}\ll 1$ that defines
the region where we want to study the physics of the model, we can
reabsorb the divergence $-\log(\eta)$ as a renormalization into it and
introduce the physical ratio $\epsilon=\frac{2}{\pi}\frac{a_B}{\lambda_L}$.
This means that we have two bare constants $\epsilon_0$ and $\eta$ and this
goes to zero through the relation $\eta=e^{-\frac{\epsilon}{\epsilon_0}}$
with $\epsilon$ the physical ratio. Then, the limit $\epsilon_0\rightarrow 0$
implies $\eta\rightarrow 0$.
Through this redefinition, the dipolar terms are
\begin{equation}
    v_k^{dip}({\bf x})=
    -\epsilon\frac{Ze^2}{\lambda_L}(-1)^n(2n+1)\frac{x}{a_B}
    \label{eq:vt2}
\end{equation}
In this way, we can now prove that the
electron in a Coulomb field and an intense laser field, such to have
the parameter $\epsilon\ll 1$,
undergoes kicking turning the problem into one of quantum chaos
\cite{cas2}.

The question of a kicked hydrogen atom has been put forward to
explain experiments on ionization of Rydberg atoms in
an intense microwave field
through the mechanism of dynamical localization
\cite{cas1,cas2}. Pioneering
work has also been done on strong fields and atoms,
using kicked models e.g. in Ref.\cite{blu}. But,
a rigorous proof of this behavior derived directly from
quantum electrodynamics has not been given yet. Here, we accomplish
this task by completing the analysis of the time-dependent part
of the Kramers-Henneberger Fourier components. To complete this
derivation, we point out that a periodic distribution as
$\delta_T(t)=\sum_{n=-\infty}^{+\infty}\delta(t-nT)$ that has
period $T$, can be defined through the coefficients of its
Fourier series. So, e.g. $\delta_T(t)=\frac{1}{T}
\sum_{n=-\infty}^{+\infty}e^{i2n\pi\frac{t}{T}}$ has all
constant coefficients given by $\frac{1}{T}$. To have
convergence of the Fourier series in the sense of distributions,
the coefficients $f_n$  must satisfy the criterium of being slowly varying
\cite{zem}, i.e. $|f_n|\le M|n|^k$, being $M$ and $k$ two constants.

Then, by using eqs.(\ref{eq:vt2}) and one gets the dipolar potential as
\begin{equation}
   \tilde{V}_{KH}=
   -i\frac{Ze^2}{\lambda_L}\frac{x}{a_B}
   \epsilon
   \sum_{k=-\infty}^{+\infty}(2k+1)e^{i(2k+1)\omega t}
   =2\epsilon\frac{Ze^2}{\lambda_L}\frac{x}{a_B}
   \sum_{k=0}^{+\infty}(2k+1)\sin((2k+1)\omega t).
\end{equation}
The Fourier series appearing here has convergence just in
the sense of distributions. We now give an explicit form
for the dipolar term. Indeed, one has
\begin{equation}
    \tilde{\delta}_{\frac{T}{2}}(t)=
    \sum_{k=-\infty}^{+\infty}(-1)^k\delta\left(t-k\frac{T}{2}\right)=
    \frac{2}{T}\sum_{k=-\infty}^{+\infty}e^{i(2k+1)\omega t}
\end{equation}
so that, the dipolar term is defined through the derivative of a
periodic distribution of period $\frac{T}{2}$. One can write at last
\begin{equation}
   \tilde{V}_{KH}=
   -\epsilon\frac{\pi}{\omega^2}\frac{Ze^2}{\lambda_L}
   \frac{x}{a_B}
   \frac{d}{dt}\tilde{\delta}_{\frac{T}{2}}(t)+\cdots
\end{equation}
and we draw the conclusion that the motion of an atom in an intense
laser field undergoes kicking with period $\frac{T}{2}$ so that,
localization, a typical effect of quantum chaos, can happen.

Now, we show that the Rydberg atoms used in ionization experiments
with microwave are indeed in a regime where the ratio between the
atomic dimensions and the amplitude of the
quiver motion of the free-electron $\lambda_L$ in the laser field
is largely smaller than unity,
and the above theory applies. As a starting
point we take Ref.\cite{cas3} for a typical experiment. The less
favourable case is for an intensity of the microwave field of
2.5 V/cm, at 12.4 GHz and the Rydberg atom has $n_0$=98. This
gives for the ratio
$\frac{n_0^2a_B}{\lambda_L}=\frac{n_0^2\omega}{\sqrt{8I_BU_p}}$ the value
0.0027, largely lesser than unity. Instead, the most favourable
case is given by the intensity of the electric field of 21 V/cm,
at 18 Ghz and $n_0$=64 giving a ratio of 0.00029 improving the
situation of about a magnitude order. Then, we can conclude that
Rydberg atoms are kicked by the microwave field and localization
can happen, in agreement with all the current researches. This
result comes directly from quantum electrodynamics and so, it
is fully justified.

For the case of the harmonic generation the situation is still
better because the Keldysh parameter $\gamma$ given by
$\sqrt{\frac{I_B}{2U_p}}$ can be lesser than one permitting
a straightforward application of the perturbation theory. A
typical example of an experiment in this regime is given in
Ref.\cite{hb}. This possibility can be fully exploited if
a set of unperturbed states can be found to do perturbation
theory. This is indeed the case as we will see in the next
section.

\subsection{Rigidity of the Atomic Wave Function}

In order to do perturbation theory, one generally needs a
full set of orthonormal functions to start with, representing
the unperturbed system. But, here one has to diagonalize the
Hamiltonian
\begin{equation}
    H_0=\frac{{\bf p}^2}{2m}+\int_{-1}^1\frac{dx'}{\pi}
    V({\bf x}-\mbox{\boldmath $\epsilon$}\lambda_L x')
    \frac{1}{\sqrt{1-x'^2}} \label{eq:h0}
\end{equation}
that is generally an impossible task analitically.

What we want to do is to analyze this problem in the limit
where the amplitude of the quiver motion of the free-electron
in the laser field is much larger than the atomic radius,
further specializing the above problem to the Coulomb potential.

To evaluate the degree of deformation due to the laser field in this
approximation we apply
a modified Rayleigh-Schr\"odinger approximation
as obtainable from the time-dependent perturbation series.
This can be accomplished easily if one makes the multipolar
expansion of the dressed Coulomb potential in eq.(\ref{eq:h0}) as
\begin{equation}
    V_{KH}=-\frac{Ze^2}{r}\left[1+
    \sum_{n=1}^{+\infty}A_n\left(\frac{\lambda_L}{r}\right)^{2n}
    P_{2n}\left(\frac{x}{r}\right)\right] \label{eq:dp}
\end{equation}
being $A_n=\int_{-1}^1dxx^{2n}/(\pi\sqrt{1-x^2})$ and $P_n$ the $n$-th
Legendre polynomial. So, we approach the static part of the
Kramers-Henneberg Hamiltonian differently from the time-dependent
part. The reason to do that is that we expect very large shifts
of the energy levels and that just very few terms of the multipolar
series (\ref{eq:dp}) really contributes to the matrix elements in
the Rayleigh-Schr\"odinger series.

The way we compute the Rayleigh-Schr\"odinger corrections is taken
from the time-dependent perturbation theory. Indeed, let us suppose
that one can neglect the time dependent part of the Kramers-Henneberger
Hamiltonian, so that one is left with the time-independent problem
\begin{equation}
    H_A=\frac{{\bf p}^2}{2m}-\frac{Ze^2}{r}+\delta V_{KH}({\bf x})
\end{equation}
where we have put
\begin{equation}
    \delta V_{KH}=-\frac{Ze^2}{r}
    \sum_{n=1}^{+\infty}A_n\left(\frac{\lambda_L}{r}\right)^{2n}
    P_{2n}\left(\frac{x}{r}\right)
\end{equation}
and we want to solve the problem
\begin{equation}
    H_A|\psi(t)\rangle=i\frac{\partial |\psi(t)\rangle}{\partial t}.
\end{equation}
This is the way stabilization is studied in the Kramers-Henneberger
frame when the limit $\omega\rightarrow\infty$ is taken\cite{kk1,sta}.
Here we are able to prove that stabilization indeed exists in the limit
of large ratio of free-electron quiver motion and Bohr radius due
to the rigidity of the atomic wave function.

We take as unperturbed states the ones of the Coulomb problem by
setting
\begin{equation}
    |\psi(t)\rangle=\sum_n a_n(t) e^{-iE_n t}|n\rangle
\end{equation}
being
\begin{equation}
    \left(\frac{{\bf p}^2}{2m}-\frac{Ze^2}{r}
    \right)|n\rangle=E_n|n\rangle.
\end{equation}
One gets for the amplitudes
\begin{equation}
    ia_m(t)=\langle m|\delta V_{KH}({\bf x})|m\rangle a_m(t)+
    \sum_{n\ne m} e^{-i(E_n-E_m)t}
    \langle m|\delta V_{KH}({\bf x})|n\rangle a_n(t)
\end{equation}
and introducing $b_m(t)=e^{-i\delta E_m t}a_m(t)$,
$\delta E_m=\langle m|\delta V_{KH}({\bf x})|m\rangle$ and
$\tilde E_m=E_m+\delta E_m$, we arrive finally at the equations
\begin{equation}
    ib_m(t)=
    \sum_{n\ne m} e^{-i(\tilde E_n-\tilde E_m)t}
    \langle m|\delta V_{KH}({\bf x})|n\rangle b_n(t).
\end{equation}
Then, if the shifts $\delta E_m$ are really large
in the limit of large ratio between the quiver
amplitude of the motion of the free electron and
the Bohr radius, the Rayleigh-Schr\"odinger corrections
to the initial wave-function are really small and this
state is ``rigid'' with respect to the perturbation
introduced by the laser field. Indeed, one has
\begin{equation}
    b_m(t)=b_m(0)+
    \sum_{n\ne m} e^{-i(\tilde E_n-\tilde E_m)t}
    \frac{\langle m|\delta V_{KH}({\bf x})|n\rangle}
    {\tilde E_n-\tilde E_m}b_n(0)+\cdots
\end{equation}
where an adiabatic switching of the perturbation has been introduced.
We see that the first order correction
gives a modification of the Rayleigh-Schr\"odinger perturbation
theory as in place of the unperturbed energy levels $E_m$ there
are the modified energy levels $\tilde E_m$. This appears also as
an improvement with respect to the Brillouin-Wigner perturbation
series. Indeed, one has for the wave function
\begin{equation}
    |\psi(t)\rangle=\sum_ne^{-i\tilde E_nt}a_n(0)|n\rangle+
    \sum_ne^{-i\tilde E_nt}\sum_{k\ne n}
    e^{-i(\tilde E_k-\tilde E_n)t}
    \frac{\langle n|\delta V_{KH}({\bf x})|k\rangle}
    {\tilde E_k-\tilde E_n}a_k(0)|k\rangle + \cdots. \label{eq:ser}
\end{equation}
In this way is possible to show that the
first-order correction is really small.

For the level shift one gets for the first few levels (we have
put $a_B$ for the Bohr radius, $l$ for the orbital angular
moment and $l_z$ for the third component of the orbital
angular moment):

n=1

\begin{eqnarray}
    \delta E_{001}&=&
    \langle l=0,l_z=0,n=1|\delta V_{KH}({\bf x})|l=0,l_z=0,n=1\rangle=0
\end{eqnarray}

n=2

\begin{eqnarray}
    \delta E_{102}&=&
    \langle l=1,l_z=0,n=2|\delta V_{KH}({\bf x})|l=1,l_z=0,n=2\rangle
    =\frac{1}{240}\frac{Ze^2}{a_B}\left(\frac{\lambda_L}{a_B}\right)^2
    \nonumber \\
    \delta E_{112}&=&\delta E_{1-12}=
    \langle l=1,l_z=1,n=2|\delta V_{KH}({\bf x})|l=1,l_z=1,n=2\rangle
    =-\frac{1}{480}\frac{Ze^2}{a_B}\left(\frac{\lambda_L}{a_B}\right)^2
    \nonumber \\
    \delta E_{002}&=&
    \langle l=0,l_z=0,n=2|\delta V_{KH}({\bf x})|l=0,l_z=0,n=2\rangle
    =0
\end{eqnarray}

n=3

\begin{eqnarray}
    \delta E_{223}&=&
    \langle l=2,l_z=2,n=3|\delta V_{KH}({\bf x})|l=2,l_z=2,n=3\rangle
    =-\frac{Ze^2}{a_B}
    \left(\frac{1}{5760}\left(\frac{\lambda_L}{a_B}\right)^2
    +\frac{1}{816480}\left(\frac{\lambda_L}{a_B}\right)^4\right)
    \nonumber \\
    \delta E_{213}&=&\delta E_{2-13}=
    \langle l=2,l_z=1,n=3|\delta V_{KH}({\bf x})|l=2,l_z=1,n=3\rangle
    =\frac{Ze^2}{a_B}
    \left(\frac{1}{11340}\left(\frac{\lambda_L}{a_B}\right)^2
    +\frac{1}{204120}\left(\frac{\lambda_L}{a_B}\right)^4\right)
    \nonumber \\
    \delta E_{203}&=&
    \langle l=2,l_z=0,n=3|\delta V_{KH}({\bf x})|l=2,l_z=0,n=3\rangle
    =-\frac{Ze^2}{a_B}
    \left(-\frac{1}{5670}\left(\frac{\lambda_L}{a_B}\right)^2
    +\frac{1}{136080}\left(\frac{\lambda_L}{a_B}\right)^4\right)
    \nonumber \\
    \delta E_{103}&=&
    \langle l=1,l_z=0,n=3|\delta V_{KH}({\bf x})|l=1,l_z=0,n=3\rangle
    =\frac{1}{810}\frac{Ze^2}{a_B}\left(\frac{\lambda_L}{a_B}\right)^2
    \nonumber \\
    \delta E_{113}&=&\delta E_{1-13}=
    \langle l=1,l_z=1,n=3|\delta V_{KH}({\bf x})|l=1,l_z=1,n=3\rangle
    =-\frac{1}{1620}\frac{Ze^2}{a_B}\left(\frac{\lambda_L}{a_B}\right)^2
    \nonumber \\
    \delta E_{003}&=&
    \langle l=0,l_z=0,n=3|\delta V_{KH}({\bf x})|l=0,l_z=0,n=3\rangle
    =0
\end{eqnarray}

n=4

\begin{eqnarray}
    \delta E_{334}&=&
    \langle l=3,l_z=3,n=4|\delta V_{KH}({\bf x})|l=3,l_z=3,n=4\rangle
    \nonumber \\
    &=&-\frac{Ze^2}{a_B}
    \left(\frac{1}{32256}\left(\frac{\lambda_L}{a_B}\right)^2
    +\frac{3}{50462720}\left(\frac{\lambda_L}{a_B}\right)^4
    +\frac{25}{113359454208}\left(\frac{\lambda_L}{a_B}\right)^6
    \right) \nonumber \\
    \delta E_{324}&=&\delta E_{3-24}=
    \langle l=3,l_z=2,n=4|\delta V_{KH}({\bf x})|l=3,l_z=2,n=4\rangle
    \nonumber \\
    &=&-\frac{Ze^2}{a_B}
    \left(\frac{1}{7208960}\left(\frac{\lambda_L}{a_B}\right)^4
    +\frac{25}{18893242368}\left(\frac{\lambda_L}{a_B}\right)^6
    \right) \nonumber \\
    \delta E_{314}&=&\delta E_{3-14}=
    \langle l=3,l_z=1,n=4|\delta V_{KH}({\bf x})|l=3,l_z=1,n=4\rangle
    \nonumber \\
    &=&-\frac{Ze^2}{a_B}
    \left(-\frac{1}{53760}\left(\frac{\lambda_L}{a_B}\right)^2
    +\frac{1}{50462720}\left(\frac{\lambda_L}{a_B}\right)^4
    +\frac{125}{37786484736}\left(\frac{\lambda_L}{a_B}\right)^6
    \right) \nonumber \\
    \delta E_{304}&=&
    \langle l=3,l_z=0,n=4|\delta V_{KH}({\bf x})|l=3,l_z=0,n=4\rangle
    \nonumber \\
    &=&-\frac{Ze^2}{a_B}
    \left(-\frac{1}{40320}\left(\frac{\lambda_L}{a_B}\right)^2
    +\frac{3}{25231360}\left(\frac{\lambda_L}{a_B}\right)^4
    -\frac{125}{28339863552}\left(\frac{\lambda_L}{a_B}\right)^6
    \right) \nonumber \\
    \delta E_{224}&=&\delta E_{2-24}=
    \langle l=2,l_z=2,n=4|\delta V_{KH}({\bf x})|l=2,l_z=2,n=4\rangle
    \nonumber \\
    &=&-\frac{Ze^2}{a_B}
    \left(\frac{1}{13440}\left(\frac{\lambda_L}{a_B}\right)^2
    +\frac{3}{4587520}\left(\frac{\lambda_L}{a_B}\right)^4\right)
    \nonumber \\
    \delta E_{214}&=&\delta E_{2-14}=
    \langle l=2,l_z=1,n=4|\delta V_{KH}({\bf x})|l=2,l_z=1,n=4\rangle
    \nonumber \\
    &=&\frac{Ze^2}{a_B}
    \left(\frac{1}{26880}\left(\frac{\lambda_L}{a_B}\right)^2
    +\frac{3}{1146880}\left(\frac{\lambda_L}{a_B}\right)^4\right)
    \nonumber \\
    \delta E_{204}&=&
    \langle l=2,l_z=0,n=4|\delta V_{KH}({\bf x})|l=2,l_z=0,n=4\rangle
    \nonumber \\
    &=&-\frac{Ze^2}{a_B}
    \left(-\frac{1}{13440}\left(\frac{\lambda_L}{a_B}\right)^2
    +\frac{9}{2293760}\left(\frac{\lambda_L}{a_B}\right)^4\right)
    \nonumber \\
    \delta E_{114}&=&\delta E_{1-14}=
    \langle l=1,l_z=1,n=4|\delta V_{KH}({\bf x})|l=1,l_z=1,n=4\rangle
    =-\frac{1}{3840}\frac{Ze^2}{a_B}\left(\frac{\lambda_L}{a_B}\right)^2
    \nonumber \\
    \delta E_{104}&=&
    \langle l=1,l_z=0,n=4|\delta V_{KH}({\bf x})|l=1,l_z=0,n=2\rangle
    =\frac{1}{1920}\frac{Ze^2}{a_B}\left(\frac{\lambda_L}{a_B}\right)^2
    \nonumber \\
    \delta E_{004}&=&
    \langle l=0,l_z=0,n=4|\delta V_{KH}({\bf x})|l=0,l_z=0,n=4\rangle
    =0
\end{eqnarray}
It easily realized that just very few terms of the multipolar
series of $\delta V_{KH}$ really contribute to level shifts making
the argument working. Beside, as it should be expected the s-states
have no shift.

Indeed, for the matrix elements, assuming as initial state the
ground state of the atom one obtains

n=2

\begin{eqnarray}
    \langle l=1,l_z=1,n=2|\delta V_{KH}({\bf x})|l=0,l_z=0,n=1\rangle&=&0
    \nonumber \\
    \langle l=1,l_z=0,n=2|\delta V_{KH}({\bf x})|l=0,l_z=0,n=1\rangle&=&0
    \nonumber \\
    \langle l=0,l_z=0,n=2|\delta V_{KH}({\bf x})|l=0,l_z=0,n=1\rangle&=&0
\end{eqnarray}

n=3

\begin{eqnarray}
    \langle l=2,l_z=2,n=3|\delta V_{KH}({\bf x})|l=0,l_z=0,n=1\rangle&=&
    -\frac{Ze^2}{a_B}\frac{1}{720}\left(\frac{\lambda_L}{a_B}\right)^2
    \nonumber \\
    \langle l=2,l_z=1,n=3|\delta V_{KH}({\bf x})|l=0,l_z=0,n=1\rangle&=&0
    \nonumber \\
    \langle l=2,l_z=0,n=3|\delta V_{KH}({\bf x})|l=0,l_z=0,n=1\rangle&=&
    \frac{Ze^2}{a_B}\frac{\sqrt{150}}{10800}
    \left(\frac{\lambda_L}{a_B}\right)^2
    \nonumber \\
    \langle l=1,l_z=1,n=3|\delta V_{KH}({\bf x})|l=0,l_z=0,n=1\rangle&=&0
    \nonumber \\
    \langle l=1,l_z=0,n=3|\delta V_{KH}({\bf x})|l=0,l_z=0,n=1\rangle&=&0
    \nonumber \\
    \langle l=0,l_z=0,n=3|\delta V_{KH}({\bf x})|l=0,l_z=0,n=1\rangle&=&0
\end{eqnarray}

n=4

\begin{eqnarray}
    \langle l=3,l_z=3,n=4|\delta V_{KH}({\bf x})|l=0,l_z=0,n=1\rangle&=&0
    \nonumber \\
    \langle l=3,l_z=2,n=4|\delta V_{KH}({\bf x})|l=0,l_z=0,n=1\rangle&=&0
    \nonumber \\
    \langle l=3,l_z=1,n=4|\delta V_{KH}({\bf x})|l=0,l_z=0,n=1\rangle&=&0
    \nonumber \\
    \langle l=3,l_z=0,n=4|\delta V_{KH}({\bf x})|l=0,l_z=0,n=1\rangle&=&0
    \nonumber \\
    \langle l=2,l_z=2,n=4|\delta V_{KH}({\bf x})|l=0,l_z=0,n=1\rangle&=&
    -\frac{Ze^2}{a_B}\frac{13\sqrt{150}}{150000}
    \left(\frac{\lambda_L}{a_B}\right)^2
    \nonumber \\
    \langle l=2,l_z=1,n=4|\delta V_{KH}({\bf x})|l=0,l_z=0,n=1\rangle&=&0
    \nonumber \\
    \langle l=2,l_z=0,n=4|\delta V_{KH}({\bf x})|l=0,l_z=0,n=1\rangle&=&
    \frac{Ze^2}{a_B}\frac{13}{15000}\left(\frac{\lambda_L}{a_B}\right)^2
    \nonumber \\
    \langle l=1,l_z=1,n=4|\delta V_{KH}({\bf x})|l=0,l_z=0,n=1\rangle&=&0
    \nonumber \\
    \langle l=1,l_z=0,n=4|\delta V_{KH}({\bf x})|l=0,l_z=0,n=1\rangle&=&0
    \nonumber \\
    \langle l=0,l_z=0,n=4|\delta V_{KH}({\bf x})|l=0,l_z=0,n=1\rangle&=&0.
\end{eqnarray}
Again we see that just very few terms of $\delta V_{KH}$ really
contribute to the matrix elements. Beside, this contribution is
smaller if not zero with respect to the level shifts. Then,
these results strongly support
the statement that the first-order correction into the
series (\ref{eq:ser}) is really small in the limit $\frac{\lambda_L}{a_B}
\gg 1$ as it should be. This also supports stabilization when the limit
of laser frequency going to infinity is taken, with the ponderomotive
energy being constant as in this limit, the time-dependent part of the
Kramers-Henneberger Hamiltonian can be neglected \cite{kk1,sta}.

Finally, we conclude that a fairly good approximation for the
unperturbed Hamiltonian for our aims is
\begin{equation}
    H_A\approx\sum_n\tilde E_n|n\rangle\langle n| \label{eq:dia}
\end{equation}
where just diagonal terms are kept and the off-diagonal terms
\begin{equation}
    H'_A=\sum_{n\neq m}|m\rangle\langle n|\langle m|\delta V_{KH}|n\rangle
\end{equation}
are neglected.
We take the Hamiltonian (\ref{eq:dia}) for
time-dependent computations in perturbation theory.

\section{Perturbation Theory for Atoms in Intense Laser Fields}
\label{per}

The experiments carried out to produce harmonics have a small
laser frequency with respect to ionization and ponderomotive energy,
then also the time dependent components of the Kramers-Henneberger
Hamiltonian need to be considered.

Our aim is to make a perturbaion analysis of the Hamiltonian
\begin{equation}
    H=\sum_n\tilde E_n|n\rangle\langle n|
    +2\epsilon\omega\gamma\frac{x}{a_B}
   \sum_{k=0}^{+\infty}(2k+1)\sin((2k+1)\omega t)
\end{equation}
being
\begin{equation}
    \gamma=\frac{Ze^2}{\lambda_L\omega}=\sqrt{\frac{I_B}{2U_p}}
\end{equation}
the Keldysh parameter. So, the perturbation theory is applicable
in this case just when $\gamma\ll 1$ that defines the so-called
tunnelling regime for the electron in the laser field. But, the
situation here is more favourable as we have just to require
that the distance between energy levels
is much larger than $\omega\gamma$ and we
are able to account for a larger number of experiments than the
tunnelling regime would permit.

We write the equations for the probability amplitudes as
\begin{equation}
    i\dot a_m(t)=-i\epsilon\omega\gamma\sum_n
    e^{-i(\tilde E_n-\tilde E_m)t}a_n(t)
    \left\langle m\left|\frac{x}{a_B}\right|n\right\rangle
    \sum_{k=0}^{+\infty}\left[(2k+1)\left(e^{i(2k+1)\omega t}
    -e^{-i(2k+1)\omega t}\right)\right].
\end{equation}
Then, a generic term, out of resonance, will be written as
\begin{eqnarray}
    a_m(t)&=&a_m(0)+ \nonumber \\
    & &i\epsilon\omega\gamma\sum_na_n(0)
    \left\langle m\left|\frac{x}{a_B}\right|n\right\rangle
    \sum_{k=0}^{+\infty}\left[(2k+1)\left(
    \frac{e^{i((2k+1)\omega-\tilde E_n+\tilde E_m) t}-1}
    {(2k+1)\omega-\tilde E_n+\tilde E_m}+\right.\right. \nonumber \\
    & &\left.\left.\frac{e^{-i((2k+1)\omega+\tilde E_n-\tilde E_m) t}-1}
    {(2k+1)\omega+\tilde E_n-\tilde E_m}
    \right)\right]+\cdots. \label{eq:amp2}
\end{eqnarray}
This result shows that when there are
two resonant states $m$ and $n$ in the laser
field, then we have Rabi flopping with Rabi frequency given by
\begin{equation}
    \frac{\Omega_R}{2}=
    \epsilon\gamma(2k+1)\omega
    \left|\left\langle m\left|\frac{x}{a_B}\right|n\right\rangle\right|
\end{equation}
determined by the $(2k+1)$-th harmonic of the laser frequency.
This implies a significant modification of the spectrum of the
harmonics with respect to the observed patterns.
Otherwise, out of resonance amplitudes are really small due to
the large shifts in the limit of large $\frac{\lambda_L}{a_B}$ ratio.
So, the interesting case is the resonance with the
set of continuous states of the atom representing the
situation of the recollision model. In this case, instead of
flopping, we can have decay, that means ionization,
and a probability for the electron
to turn back to the core emitting radiation.

Then, we can apply the Wigner-Weisskopf argument \cite{ww} to
eq.(\ref{eq:amp2}). The first approximation is to reduce the
system to the two levels really resonating and to assume all
the other amplitudes to be small in the sense given above.
This means to approximate eq.(\ref{eq:amp2}) by
\begin{eqnarray}
    \dot a_0(t)&\approx&-\epsilon\omega\gamma\sum_{\bf p}
    e^{-i(E_{\bf p}-E_0)t}
    \left\langle 0\left|\frac{x}
    {a_B}\right|{\bf p}\right\rangle
    a_{\bf p}(t)\sum_{k=0}^{+\infty}(2k+1)e^{i(2k+1)\omega t} \nonumber \\
    \dot a_{\bf p}(t)&\approx&\epsilon\omega\gamma
    e^{-i(E_0-E_{\bf p})t}
    \left\langle {\bf p}\left|\frac{x}
    {a_B}\right|0\right\rangle
    a_0(t)\sum_{k=0}^{+\infty}(2k+1)e^{-i(2k+1)\omega t} \label{eq:ww}
\end{eqnarray}
where we have labelled the eigenstates of continuum for the atom by the
momentum ${\bf p}$ and $E_0=-I_B$.
We seek a solution of this equations by setting
\begin{equation}
    a_0(t)=e^{-\frac{\Gamma'}{2}t}. \label{eq:dec}
\end{equation}
By substituting eq.(\ref{eq:dec})
into eq.(\ref{eq:ww}) one arrives at the condition
\begin{equation}
    \frac{\Gamma'}{2}=i\epsilon^2\gamma^2\omega^2
    \sum_{\bf p}
    \left|\left\langle 0\left|\frac{x}
    {a_B}\right|{\bf p}\right\rangle\right|^2
    \sum_{k=0}^{+\infty}(2k+1)^2
    \frac{1-e^{-i[E_{\bf p}-E_0-(2k+1)\omega]t+i\frac{\Gamma'}{2}t}}
    {(2k+1)\omega-E_{\bf p}+E_0-i\frac{\Gamma'}{2}}
\end{equation}
For small $\Gamma'$ this indeed reduces to a time-independent expression
\cite{heit}
\begin{equation}
    \frac{\Gamma'}{2}=i\epsilon^2\gamma^2\omega^2
    \sum_{\bf p}
    \left|\left\langle 0\left|\frac{x}
    {a_B}\right|{\bf p}\right\rangle\right|^2
    \sum_{k=0}^{+\infty}(2k+1)^2\left[
    {\cal P}\frac{1}
    {(2k+1)\omega-E_{\bf p}+E_0}
    -i\pi\delta(E_{\bf p}-E_0-(2k+1)\omega)\right]
\end{equation}
with ${\cal P}$ meaning the principal value. So, finally, one gets
the a.c. Stark shift of the ground state of the atom given by
\begin{equation}
    \frac{\delta\omega}{2}=\epsilon^2\gamma^2\omega^2
    \sum_{\bf p}
    \left|\left\langle 0\left|\frac{x}
    {a_B}\right|{\bf p}\right\rangle\right|^2
    \sum_{k=0}^{+\infty}(2k+1)^2
    {\cal P}\frac{1}
    {(2k+1)\omega-E_{\bf p}+E_0}
\end{equation}
and the decay rate for the ionization of the atom
\begin{equation}
    \Gamma=2\pi\epsilon^2\gamma^2\omega^2
    \sum_{\bf p}
    \left|\left\langle 0\left|\frac{x}
    {a_B}\right|{\bf p}\right\rangle\right|^2
    \sum_{k=0}^{+\infty}(2k+1)^2\delta(E_{\bf p}-E_0-(2k+1)\omega).
\end{equation}
Then, for the continuum one has
\begin{equation}
    a_{\bf p}(t)\approx -i\epsilon\gamma\omega
    \sum_{k=0}^{+\infty}
    \left\langle{\bf p}\left|\frac{x}
    {a_B}\right|0\right\rangle
    (2k+1)\frac{e^{i[E_{\bf p}-E_0-(2k+1)\omega+i\frac{\Gamma'}{2}]t}-1}
    {E_{\bf p}-E_0-(2k+1)\omega+i\frac{\Gamma'}{2}} \label{eq:ap}
\end{equation}
so that, taking
\begin{equation}
    |\psi(t)\rangle=a_0(t)e^{iI_Bt}|0>
    +\sum_{\bf p}e^{-iE_{\bf p}t}a_{\bf p}(t)|{\bf p}\rangle
\end{equation}
one has for the harmonic spectrum
\begin{equation}
    \langle\psi(t)|x|\psi(t)\rangle=
    \sum_{\bf p}\left(e^{i(E_{\bf p}+I_B)t}a_0(t)a_{\bf p}^*(t)
    \left\langle{\bf p}\left|x\right|0\right\rangle
    +e^{-i(E_{\bf p}+I_B)t}a_{\bf p}(t)a_0^*(t)
    \left\langle 0\left|x\right|{\bf p}\right\rangle\right)
\end{equation}
where use has been made of the fact that $\langle 0|x|0\rangle=
\langle{\bf p}|x|{\bf p}\rangle=0$. Assuming the laser light
coming from the far past, eq.(\ref{eq:ap}) can rewritten as
\begin{equation}
    a_{\bf p}(t)\approx -i\epsilon\gamma\omega
    \sum_{k=0}^{+\infty}
    \left\langle{\bf p}\left|\frac{x}
    {a_B}\right|0\right\rangle
    (2k+1)\frac{e^{i[E_{\bf p}-E_0-(2k+1)\omega+i\frac{\Gamma'}{2}]|t|}}
    {E_{\bf p}-E_0-(2k+1)\omega+i\frac{\Gamma'}{2}}
\end{equation}
and we are left with the resonant part of the spectrum given by
\begin{eqnarray}
    \langle\psi(t)|x|\psi(t)\rangle_R&\approx&
    -2\pi\epsilon\gamma\omega\sum_{\bf p}
    \left|
    \left\langle 0\left|\frac{x}{a_B}\right|{\bf p}\right\rangle
    \right|^2
    \sum_{n=0}^\infty(2n+1)
    \frac{\frac{\Gamma}{2\pi}}
    {\left[E_{\bf p}-E_0-(2k+1)\omega-\frac{\delta\omega}{2}\right]^2
    + \frac{\Gamma^2}{4}}\times \nonumber \\
    & &e^{-\Gamma |t|}\cos\left[(2n+1)\omega t+\frac{\delta\omega}{2}t\right].
\end{eqnarray}
This form of the spectrum gives two
main results: Firstly, one has all the harmonics shifted by the same
quantity $\frac{\delta\omega}{2}$ and secondly each harmonic has a
Lorentzian form that, for very small $\Gamma$ can be reduced to a
Dirac $\delta$ distribution.
In this case the integration in ${\bf p}$
can be done taking a plane wave as a final state
and neglecting the shift of the spectrum,
one gets an improved version of the result of Ref.\cite{fra0}, that is
\begin{equation}
    \langle\psi(t)|x|\psi(t)\rangle_R\approx\frac{2}{\pi}
    \frac{2^\frac{17}{2}}{3^\frac{9}{2}}\frac{Ze^2\omega}{U_p^2}
    \gamma^5
    \sum_{n=n_0}^{+\infty}
    \frac{x_n^\frac{3}{2}}{\left(x_n+\frac{2\gamma^2}{3}\right)^5}
    \cos((2n+1)\omega t)e^{-\Gamma |t|}.
\end{equation}
being $x_n=\frac{(2n+1)\omega-I_B}{3U_p}$ and $n_0$ is a lower cut-off
given by the integer value such that $(2n_0+1)\omega-I_B\ge 0$ firstly
foreseen in Ref.\cite{fra0}. Being scaled in this way
through $U_p$, $I_B$ and $\omega$, a simple model of a single electron
in an atom with atomic number
$Z_{eff}$ can be adopted for complex atoms in a strong
laser field and the above result used also in this case. As shown
in Ref.\cite{fra0}, the order of magnitude is correct. The same
is true for $\Gamma$ given by
\begin{equation}
    \Gamma=
    \frac{256}{3\pi^2}\frac{\omega^2}{U_p}\gamma^2
    \sum_{n=n_0}^{+\infty}
    \left[\frac{I_B}{(2n+1)\omega}\right]^\frac{5}{2}
    \left[1-\frac{I_B}{(2n+1)\omega}\right]^\frac{3}{2}
\end{equation}
that for the experiment described in Ref.\cite{hb} yields for helium
.012 eV and for neon .01 eV improving the results given in Ref.\cite{fra0}.
This in turn means that the mean lifetime of the
harmonics ranges from 52 to 68 fs and, by e-folding, after a time of
few 10$^2$ fs the harmonics disappear in agreement with the
experimental results. The result is independent from the order of
the harmonics. We expect that the value of $\Gamma$ should be in
better agreement with the experimental results when $U_p$ increases
or being the same, when the Keldysh parameter $\gamma$ decreases marking the
so called tunnelling regime.

\section{Conclusions}

We have given a full theory of harmonic generation starting from
non-relativistic quantum electrodynamics in the dipole
approximation. A relation with quantum chaos is obtained
through a kicked model that originates from the model at
small distances where a regularization procedure is
needed. Perturbation theory can be done in the tunnelling
regime, where the Keldysh parameter is small, as the
atomic wavefunction turns out to be rigid due to the
way the energy levels are shifted by the laser field. In
this way a deep connection between ionization rate and
harmonic spectrum is given. Finally, a fully understanding of the
recollision model is obtained and this was our main aim.

\label{end}


\begin{references}
\bibitem{kk1} M. Protopapas, C. H. Keitel and P. L. Knight,
Rep. Prog. Phys. {\bf 60}, 389 (1997) and refs. therein.
\bibitem{Cork} J. L. Krause, K. J. Schafer and K. C. Kulander,
Phys. Rev. Lett. {\bf 68}, 3535 (1992); Phys. Rev. A {\bf 45},
4998 (1992); P. B. Corkum, Phys. Rev. Lett. {\bf 71}, 1994 (1993).
\bibitem{kk2} M. Protopapas, D. G. Lappas, C. H. Keitel and P. L. Knight,
Phys. Rev. A {\bf 53}, R2933 (1996)
\bibitem{Huil} M. Lewenstein, P. Balcou, M. Yu. Ivanov, A. L'Huillier
and P. Corkum, Phys. Rev. A {\bf 49}, 2117 (1994).
\bibitem{kul} K. C. Kulander, K. J. Schafer and J. L. Krause,
{\sl Atoms in Intense Radiation Fields}, edited by M. Gavrila (Academic, New York, 1992)
\bibitem{mil} B. Sundaram and P. W. Milonni, Phys. Rev. A
{\bf 41}, 6571 (1990).
\bibitem{kei} F. I. Gauthey, C. H. Keitel, P. L. Knight and
A. Maquet, Phys. Rev. A {\bf 55}, 615 (1997).
\bibitem{maq} M. L. Pons, R. Ta\"{$\!\!\!\imath$}eb and A. Maquet, Phys. Rev.
A {\bf 54}, 3634 (1996).
\bibitem{fra0} M. Frasca, Phys. Lett. A {\bf 260}, 149 (1999).
\bibitem{fra1} M. Frasca, Phys. Rev. A {\bf 58}, 3439 (1998).
\bibitem{fra2} M. Frasca, Phys. Rev. A {\bf 60}, 573 (1999).
\bibitem{ma} R. E. O'Malley, Jr., {\sl Singular Perturbation Methods for
Ordinary Differential Equations}, (Springer-Verlag, New York, 1991); the
original proof is given in G. D. Birkhoff, Trans. Amer. Math. Soc. {\bf 9},
219 (1908).
\bibitem{scul} C. Cohen-Tannoudji, J. D. Dupont-Roc and G. Grynberg,
{\sl Photons and Atoms}, (Wiley, New York, 1989);
{\sl Atom-Photon Interaction}, (Wiley, New York, 1992).
\bibitem{kh} H.~A.~Kramers, {\sl Collected Scientific Papers},
(North-Holland, Amsterdam, 1956); W.~C.~Henneberger, Phys. Rev. Lett.,
{\bf 21}, 838 (1968).
\bibitem{cas1} G. Casati, B. V. Chirikov, D. L. Shepelyansky and I. Guarneri,
Phys. Rep. {\bf 154}, 77 (1987); G. Casati, D. L. Shepelyansky and I. Guarneri,
IEEE J. of Quantum Electr. {\bf 24}, 1420 (1988).
\bibitem{cas2} G. Casati and B. V. Chirikov (eds), {\sl Quantum Chaos},
(Cambridge University Press, Cambridge, 1995).
\bibitem{hb} A. L'Huillier and Ph. Balcou, Phys. Rev. Lett. {\bf 70}, 774 (1993).
\bibitem{kel} L. V. Keldysh, Sov. Phys. JETP {\bf 20}, 1307 (1965);
F. H. M. Faisal, J. Phys. B {\bf 6}, L89 (1973); H. R. Reiss, Phys. Rev. A {\bf 22},
1786 (1980).
\bibitem{bec} A. Lohr, M. Kleber, R. Kopold and W. Becker, Phys. Rev. A {\bf 55}, R4003 (1997);
W. Becker, A. Lohr, M. Kleber and M. Lewenstein, Phys. Rev. A {\bf 56}, 645 (1997).
\bibitem{gao} J. Gao, F. Shen and J. G. Eden, Phys. Rev. A {\bf 61}, 043812 (2000)
\bibitem{fra3} M. Frasca, to appear in J. Mod. Opt., and Refs. therein.
\bibitem{cds}F.~A.~M.~de~Oliveira, M.~S.~Kim, P.~L.~Knight and
V.~Bu\u{z}ek, Phys. Rev. A, {\bf 41}, 2645 (1990);
K.~B.~M\mbox{\o}ller,
T.~G.~J\mbox{\o}rgensen and J.~P.~Dahl, Phys. Rev. A, {\bf 54}, 5378 (1996);
M.~M.~Nieto, Phys. Lett. A {\bf 229}, 135 (1997).
\bibitem{sg} L. Susskind and J. Glogower, Physics {\bf 1}, 49 (1964).
\bibitem{pb} D. T. Pegg and S. M. Barnett, Phys. Rev. A {\bf 39},
1665 (1989).
\bibitem{hak} D. F. Walls and G. J. Milburn, {\sl Quantum Optics},
(Springer, Berlin, 1994).
\bibitem{per} P. P. Corso, L. Lo Cascio and F. Persico, Phys. Rev. A {\bf 58}, 1549 (1998).
\bibitem{flo} O. E. Alon, V. Averbukh and N. Moiseyev,
Phys. Rev. Lett. {\bf 80}, 3743 (1998).
\bibitem{fra4} M. Frasca, Phys. Rev. A {\bf 56}, 1548 (1997).
\bibitem{blu} R. Bl\"umel and U Smilansky, Phys. Rev. Lett. {\bf 52}, 137
(1984); Phys. Rev. A {\bf 30}, 1040 (1984); J. Zakrzewski and K. Zyczkowski,
Phys. Rev. A {\bf 36}, 4311 (1987).
\bibitem{zem} A. H. Zemanian, {\sl Distribution Theory and Transform Analysis},
(McGraw-Hill, New York, 1965).
\bibitem{cas3} J. D. Bayfield, G. Casati, I. Guarneri and D. W. Sokol,
Phys. Rev. Lett. {\bf 63}, 364 (1987).
\bibitem{sta} M.~V.~Fedorov, {\sl Atomic and Free Electrons in a
Strong Light Field}, (World Scientific, Singapore, 1997).
\bibitem{ww} V. F. Weisskopf and E. P. Wigner, Z. Physik {\bf 63},
54 (1930). A textbook derivation is given e.g. in E. Merzbacher,
{\sl Quantum Mechanics}, (Wiley, New York, 1970).
\bibitem{heit} W. Heitler, {\sl The Quantum Theory of Radiation},
(Oxford University Press, London, 1954).
\end{references}
\end{document}